\documentclass[%
 reprint,
 amsmath,amssymb,
 aps,
]{revtex4-1}

\usepackage{graphicx}% Include figure files
\usepackage{dcolumn}% Align table columns on decimal point
\usepackage{bm}% bold math
\begin{document}

\preprint{APS/123-QED}

\title{Real-time free-running time scale with remote clocks on fiber-based frequency network}%

\author{Y.C. Guo,$^{1}$ B. Wang,$^{1,*}$ F.M. Wang,$^1$ F.F. Shi,$^3$ A.M. Zhang,$^4$ X. Zhu,$^5$ J. Yang,$^5$ K.M. Feng,$^5$ C.H. Han,$^3$ T.C. Li,$^4$ and L.J. Wang$^{1,2}$}

\affiliation{
$^1$State Key Laboratory of Precision Measurement Technology and Instruments, Department of Precision Instrument, Tsinghua University, Beijing 100084, China\\
$^2$Department of Physics, Tsinghua University, Beijing 100084, China\\
$^3$Beijing Satellite Navigation Center, Beijing 100094, China\\
$^4$National Institute of Metrology, Beijing 100013, China\\
$^5$Beijing Institute of Radio Metrology and Measurement, Beijing 100854, China\\
$^*$Corresponding author: bo.wang@tsinghua.edu.cn}

\date{\today}% It is always \today, today,
             %  but any date may be explicitly specified

\begin{abstract}
In this paper, we propose a real-time free-running time scale based on four remote hydrogen masers. The clocks in the ensemble were scattered around Beijing, connected by urban fiber links using a novel frequency synchronization system. The remote clock ensemble prevents the time scale from potential problems caused by correlation among co-located clocks. Insofar as it is real-time, it fulfills the requirements for applications such as navigation, telecommunications and so on. The free-running time scale is updated every 1200 s, and a disturbance-resistant algorithm makes it robust to fiber link disturbances and clock malfunctions. The results of a continuous experiment over 224 days are reported. The stability of the time scale outperformed any clock in the ensemble for averaging times of more than approximately $10^4$ s. 
%\begin{description}
%%\item[Usage]
%%Secondary publications and information retrieval purposes.
%\item[PACS numbers]
%06.30.Ft, 06.20.-f, 06.90.+v
%%\item[Structure]
%%You may use the \texttt{description} environment to structure your abstract;
%%use the optional argument of the \verb+\item+ command to give the category of each item. 
%\end{description}
\end{abstract}

%\pacs{06.30.Ft, 06.20.-f, 06.90.+v}% PACS, the Physics and Astronomy
                             % Classification Scheme.
%\keywords{Suggested keywords}%Use showkeys class option if keyword
                              %display desired
\maketitle

%\tableofcontents

\section{\label{sec:level1}Introduction}

Commercial atomic clocks are often used for timekeeping that requires continuous output of time signal. Physical devices can fail, however, and this has motivated the concept of a composite time scale. A composite time scale is a synthesized clock calculated from an ensemble of atomic clocks. With an appropriate algorithm, the composite clock offers higher reliability and stability compared to any one of the atomic clocks in the ensemble~\cite{1}. 

\begin{figure}[htbp]
\includegraphics[width=\linewidth]{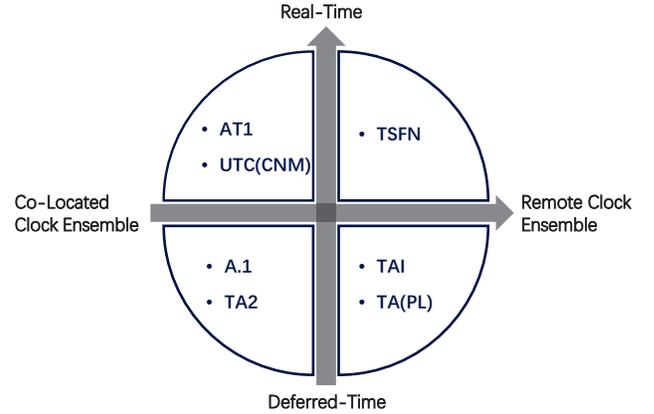}% Here is how to import EPS art
\caption{\label{fig1} Classification of composite time scales with a few examples. Time scales shown in the plot: AT1 from NIST~\cite{12,13,14}, UTC(CNM) from CENAM~\cite{15}, A.1 from USNO~\cite{22,23}, TA2 from NIST~\cite{24}, TAI from BIPM~\cite{21}, and TA(PL) from GUM~\cite{11}. The proposed TSFN is the free-running time scale based on a fiber network.}
\end{figure}

Composite time scales can be classified based on two features: if it is real-time, and if it is based on a remote clock ensemble (see Fig.~\ref{fig1}). Some applications, such as navigation, frequency metrology, and telecommunications, require a reference time scale to be real-time. However, a time scale computed in deferred time is possible to better cope with data anomalies and frequency aging. Time scales based on a co-located clock ensemble are convenient to control, monitor, and measure. However, the performance of these composites may be affected by environment-induced correlation among co-located clocks~\cite{2}. A few laboratories place H-masers in separate chambers to control their environment individually. Nevertheless, some environmental factors are too expensive or difficult to control~\cite{3}. To avoid this problem, remote clocks can be used. Additionally, time scales based on a remote clock ensemble are safer and more robust because if one institution breaks down it will not affect the continuity of the composite.

However, it is difficult to generate a real-time time scale with remote clocks. The bottleneck for this is the frequency and time synchronization between different places. Thus far, the methods in use are mainly based on satellite links~\cite{4}. Recent developments of satellite-based methods have achieved frequency transfer stability at the level of $10^{-16}$/day in terms of Allan deviation, such as carrier-phase two-way satellite frequency transfer~\cite{6}. However, this is insufficient to transfer signals of H-masers at short averaging times~\cite{7,8}. A long period of data is needed to smooth the noise induced by the transfer process. Thus, time scales based on satellite links are mostly computed in deferred time or have calculation intervals of no less than one day~\cite{4,9,10,11}.

A few time scales are listed in each category in Fig.~\ref{fig1}. Real-time time scales are mostly based on a co-located clock ensemble, e.g., the AT1 time scale of the National Institute of Standards and Technology (NIST)~\cite{12,13,14} and UTC(CNM) of the Centro Nacional de Metrologia (CENAM)~\cite{15} (Quadrant \uppercase\expandafter{\romannumeral2} in Fig.~\ref{fig1}). Some other time scales in this quadrant are introduced in ~\cite{16,17,18,19,20}. Time scales based on a remote clock ensemble are usually deferred-time, e.g., International Atomic Time (TAI) of the International Bureau of Weights and Measures (BIPM)~\cite{21} and the TA(PL) of the Central Office of Measures (GUM)~\cite{11} (Quadrant \uppercase\expandafter{\romannumeral4} in Fig.~\ref{fig1}). There are also some time scales that are co-located and computed in deferred time for better performance, e.g., the A.1 time scale of the United States Naval Observatory (USNO)~\cite{22,23} and the TA2 time scale of NIST ~\cite{24} (Quadrant \uppercase\expandafter{\romannumeral3} in Fig.~\ref{fig1}).

The first quadrant in Fig.~\ref{fig1}, representing real-time time scales based on a remote clock ensemble, is the sparsest. The tension between stability and reliability on one hand and real-time performance on the other is a long-standing problem in time scale generating. In this paper, we introduce a real-time free-running time scale based on four remote clocks in Beijing connected using fiber-based frequency synchronization technique. Hereafter, the proposed time scale is referred to as TSFN (time scale on fiber network). A disturbance-resistant algorithm is designed to ensure continuous and stable operation. Results of a 224-day continuous experiment show that the TSFN is more stable than each of the member clocks in the ensemble at averaging times from $10^4$ s to $10^6$ s.

\section{\label{sec:level1}formation of TSFN}

Recently, precise frequency synchronization methods based on fiber links have been used more widely~\cite{25,26,27,28,29,30,31,32,33,34,35}. Since 2011, fiber-based frequency synchronization systems have been developed that can reach frequency stability at a level of $10^{-15}$/s and $10^{-19}$/day in terms of Allan deviation~\cite{36}. Such a system can recover the signal of an atomic clock in a different place in real time. With this system and the abundant atomic clock resources in Beijing, we generated a regional atomic clock network based on urban fiber links, thus forming a remote atomic clock ensemble scattered across Beijing with real-time precise communication~\cite{37,38}.

\begin{figure}[htbp]
\includegraphics[width=\linewidth]{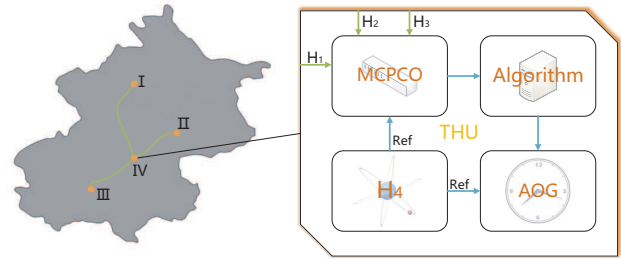}% Here is how to import EPS art
\caption{\label{fig2} Fiber-based H-maser network in Beijing and the free-running time scale system made from it. \uppercase\expandafter{\romannumeral1}, \uppercase\expandafter{\romannumeral2}, \uppercase\expandafter{\romannumeral3}, and \uppercase\expandafter{\romannumeral4}: different institutions in Beijing, of which institution \uppercase\expandafter{\romannumeral4} is Tsinghua University; $H_1$, $H_2$,  $H_3$ and $H_4$: hydrogen masers from different institutions; MCPCO: Multi-Channel Phase Comparator; THU: Tsinghua University; AOG: Frequency Standard Auxiliary Output Generator.}
\end{figure}

Currently, the network includes five hydrogen masers from four different institutions in Beijing. Two of the H-masers are located in the same room. We selected only one of these for the TSFN, to avoid the influence of correlation~\cite{2}. Fig.~\ref{fig2} shows the proposed TSFN system. $H_1$, $H_2$, and $H_3$ are the H-masers in institutions \uppercase\expandafter{\romannumeral1}, \uppercase\expandafter{\romannumeral2}, and \uppercase\expandafter{\romannumeral3}, respectively. Their signals are transferred in real time to institution \uppercase\expandafter{\romannumeral4}, which is our laboratory at Tsinghua University (THU). They are compared with the local H-maser ($H_4$) simultaneously by a Multi-Channel Phase Comparator (MCPCO). The clock data are then sent to a computer for time scale calculation. Consequently, the computer controls a Frequency Standard Auxiliary Output Generator (AOG) referenced by $H_4$ to generate a real-time free-running time scale from an ensemble of four remote H-masers. To measure the composite clock, the output of the AOG is also sent to the MCPCO to be compared with the reference clock, $H_4$.

\begin{figure}[htbp]
	\includegraphics[width=.9\linewidth]{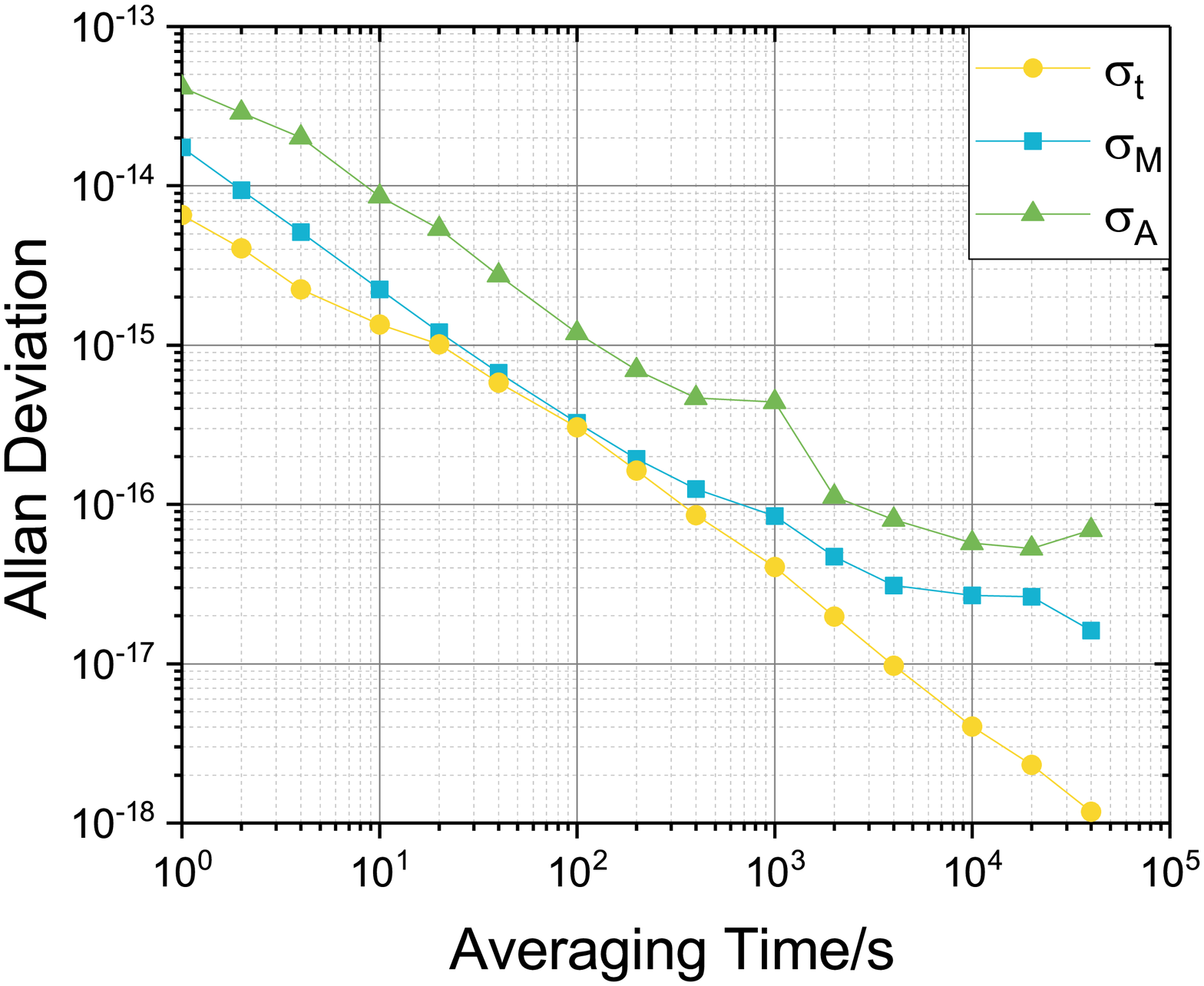}% Here is how to import EPS art
	\caption{\label{noise}System noise in the TSFN. $\sigma_t$: typical instability of the frequency transfer system; $\sigma_M$: noise floor of the MCPCO; $\sigma_A$: contribution of the AOG over the instability of its reference.}
\end{figure}

There are several main contributors to system noise in the TSFN, viz., the fiber frequency transfer systems, the MCPCO, and the AOG. Noise from each of these is shown in Fig.~\ref{noise} in terms of Allan deviation. Here, $\sigma_t$ (yellow circle) is the typical instability of the frequency transfer system, $\sigma_M$ (blue square) is the noise floor of the MCPCO, and $\sigma_A$ (green triangle) is the contribution of the AOG over the instability of its reference. The latter, i.e., $\sigma_A$, is measured by comparing the AOG output with its reference signal when the AOG is stably locked to the reference with no frequency offset. Fig.~\ref{noise} shows that the AOG contributes the most noise in the TSFN system. Indeed, the noise level from all segments is far below the stability of the H-masers in our system.

\section{\label{sec:level1}Time scale algorithm}
The algorithm for the TSFN was inspired by the algorithm for AT1 used at NIST~\cite{1,39}. It generates the composite clock iteratively. In the following, $T$ designates measured time differences, whereas $x$, $y$, and $d$ represent estimated or calculated time differences, fractional frequencies, and frequency drifts, respectively.

In each iteration, it takes $T_{ir}(t)$---the measured time offset of $H_i$ with respect to the reference clock---to estimate the time offset of the composite against the reference clock of the next iteration, marked as $\hat{x}_{cr}(t+\tau)$. Here, $\tau$ is the time interval between successive iterations. In our system, owing to the magnitude of measurement noise, as well as fractional frequencies and frequency drifts of the H-masers, the iteration interval $\tau$ is set to 1200 s. The local maser $H_4$ is used as the reference clock because, compared to other remote clocks, there is less risk of signal distortion caused by malfunctions of the transfer link.

At time epoch $t$, the algorithm estimates the time offset of each clock with respect to the composite at time epoch $t+\tau$ using the following equation:
\begin{equation}
\label{predict}
\hat{x}_{ic}(t+\tau) = x_{ic}(t)+y_{ic}(t)\cdot \tau+\frac{1}{2}d_{ic}\cdot\tau^2, i=1,2,3,4.
\end{equation}

Here, $x_{ic}(t)$ is the time offset of each clock with respect to the composite at time epoch $t$, and $y_{ic}(t)$ and $d_{ic}$ are the fractional frequency and frequency drift of $H_i$ with respect to the composite, respectively. Then, at time epoch $t+\tau$, with $T_{ir}(t+\tau)$ measured by the MCPCO, the time offset of the composite with respect to the reference clock predicted according to $H_i$ can be expressed as
\begin{equation}
\hat{x}_{cr}^{i}(t+\tau) = T_{ir}(t+\tau)-\hat{x}_{ic}(t+\tau),i=1,2,3,4.
\end{equation}

There is one set of these equations for each clock in the ensemble, including the reference clock $H_4$, where the corresponding time offset $T_{4r}(t)$ is constant for any time epoch $t$. The estimated time offset of the composite with respect to the reference clock is a weighted sum of these estimates over all the members of the ensemble:
\begin{equation}
\label{average}
\hat{x}_{cr}(t+\tau) = \sum_{i=1}^{4}\omega_i(t)\hat{x}_{cr}^{i}(t+\tau).   
\end{equation}

Then, $\hat{x}_{cr}(t+\tau)$ is used to control the AOG to obtain a real-time physical realization of the composite clock. 

The fractional frequency of each clock against the composite, $y_{ic}(t)$ in Eq. (\ref{predict}), is calculated with the following steps:
\begin{align}
f_{ic}(t) &= \frac{x_{ic}(t)-x_{ic}(t-\tau)}{\tau}-\frac{1}{2}d_{ic}\tau,\\
y_{ic}(t) &= \frac{f_{ic}(t)+\omega_y y_{ic}(t-\tau)}{1+\omega_y} + d_{ic}\tau.
\end{align}

The first step is the estimation of the average fractional frequency over the last iteration. The second is an exponential weighted average of past and present estimated frequencies. This averaging step aims to derive an unbiased frequency estimation, since with a proper iteration interval $\tau$, it is reasonable to assume that variation of the estimated frequency is mainly due to white frequency noise. Here, $\omega_y$ is a time constant to limit the number of past estimations involved in the average. Indeed, if we go back too far in time, it will be inappropriate to characterize the noise as mainly white frequency noise. This constant is set to 20 in our system, based on observations of the stability of the H-masers in the network. 

The frequency drift of clock $H_i$ with respect to the composite, $d_{ic}$ in Eq. (\ref{predict}), is considered constant during our calculation. According to the performance of each clock in a previous one-month test and with the assumption that the distribution of clock drifts is unbiased, the values of $d_{ic}$ are listed as follows:

\begin{equation}
\begin{aligned}
d_{1c} &= 9.2 \times 10^{-21}/s,\\
d_{2c} &= -1.28 \times 10^{-20}/s,\\	
d_{3c} &= -8.8 \times 10^{-21}/s,\\	
d_{4c} &= -1.8 \times 10^{-21}/s.	
\end{aligned}
\end{equation}

In Eq. (\ref{average}), $\omega_i(t)$, the weight of clock $H_i$, is computed from the average prediction error of the clock over previous iterations. The computation proceeds as follows:

\begin{align}
\varepsilon_i (t) &= \left| \hat{x}_{cr}^i (t)-\hat{x}_{cr}(t)\right| + K_i (t),\\
\label{eq8}
\sigma_i^2 (t) &= \frac{1}{N_\tau+1} \left[\varepsilon_i^2 (t)+N_\tau \sigma_i^2 (t-\tau)\right],\\
K_i (t) &= 0.5\omega_i (t-\tau) \left|\sigma_i (t-\tau)\right|,\\
\label{eq10}
\omega_i (t) &= \frac{\frac{1}{\sigma_i^2 (t)}}{\sum_{j=1}^4 \frac{1}{\sigma_j^2 (t)}}.	
\end{align}

Here, $K_i (t)$ is designed to increase the prediction error of clocks with high weights, such that it can reduce the effect of the “positive feedback loop” in weight allocation, by which highly weighted clocks tend to accumulate higher and higher weights at the cost of the others. The time constant $N_\tau$ in the exponential weighted average process in Eq. (\ref{eq8}) is set to 2 days. 

Another solution in our algorithm to account for the “positive feedback loop” is to set the maximum weight limit of each clock to 0.5. When the weight of clock $i$ calculated from Eq. (\ref{eq10}) is more than 0.5, $\omega_i (t)$ will be reset to 0.5 and the weights of the other 3 clocks will be reassigned in proportion.

\section{\label{4}Disturbance-resistant algorithm}

A basic requirement of a real-time algorithm is to provide definitive access to the time scale immediately after each measurement, with no post-processing or reprocessing. This requirement is easier to fulfill for a co-located clock ensemble than for a remote ensemble such as the TSFN. The long-distance fiber-based frequency synchronization link faces higher risk of disturbance than the short RF link used for a co-located clock ensemble. Further, as the construction of the fiber frequency network continues, new clocks will be added, and old clocks with unsatisfactory performance will be removed. A special solution is designed to ensure that the composite clock operates smoothly, regardless of fiber link disturbances or ensemble member changes.

In each iteration, we first determine the health condition of clock $H_i$ and its fiber link using a criterion $\Delta T_i (t)$, defined as
\begin{equation}
\begin{aligned}
\Delta T_i (t)=&\Big| T_{ir}(t)-\big[ T_{ir}(t-\tau)+ \\
               &\left(y_{ic}(t-\tau)- y_{rc}(t-\tau)\right)\cdot\tau\big]\Big|,  
\end{aligned}
\end{equation}
where $\Delta T_i (t)$ is the difference between the measured phase of the current iteration and the predicted phase calculated using data from previous iterations. Note that the “measured phase” is not the real phase of each clock, but rather the phase after the transfer. Thus, the health of the frequency transfer links will influence the measured phase. Two judgments are made together to decide the health of the clock:
\begin{align}
\Delta T_i (t) > C, \\
T_{ir} (t)=0,
\end{align}
where $C$ is the threshold for $\Delta T_i (t)$. In our system, we set $C$ to $1 \times 10^{-9}$ s. This was an administratively chosen constant mainly based on experience. Except for the starting point, when MCPCO measures a signal with strong or rapid variation, it will output a “zero phase”, i.e., $T_{ir}=0$.

Case 1: If either of the conditions is satisfied, there is a good chance that the fiber link or the H-maser is not working properly, or that the H-maser has been removed from the network intentionally. In this case, the measured data $T_{ir} (t)$ should not be adopted. Instead, a phase difference substitute $T'_{ir} (t)$ is used for the time scale calculation. $T'_{ir} (t)$ is calculated as
\begin{equation}
\begin{aligned}
T'_{ir} (t) =T'_{ir} (t-\tau) + \left[y_{ic} (t-\tau)- y_{rc} (t-\tau)\right]\cdot\tau.       
\end{aligned}
\end{equation}

At the beginning of the measurement, $T'_{ir}(0) = T_{ir}(0)$. This substitute ensures that TSFN operates smoothly and stably in real time even when there are strong fluctuations in the raw data. At the same time, the weight of this particular clock decreases by a step of 0.001:
\begin{equation}
\begin{aligned}
\omega_i (t)=\max\{\omega_i (t-\tau)-0.001,0\}.     
\end{aligned}
\end{equation}

The general method for determining weights is inappropriate in this case, owing to the risk of great changes to the weight assignment that can affect the stability of the composite. Thus, we control the weight such that it decreases at a preset and sufficiently slow pace. The decreasing step, 0.001, was also an administrative decision based on experience. 

Case 2: If both conditions are not satisfied, clock $H_i$ and its transfer link are considered to function well at the current iteration, which means that the measured phase difference $T_{ir} (t)$ is credible. In this case, the substitute $T'_{ir} (t)$ is calculated as follows:
\begin{equation}
\begin{aligned}
T'_{ir} (t) = T_{ir} (t)+a_i (t) .     
\end{aligned}
\end{equation}

Here, $a_i(t)$ is re-calculated only when the criterion $\Delta T_i (t)$ changes from Case 1 to Case 2, following the equation below. In other cases, it remains the same as it was in the last iteration, $a_i(t)=a_i(t-\tau)$.
\begin{equation}
\begin{aligned}
a_i(t) = T'_{ir}(t-\tau) + \left[ y_{ic}(t-\tau)-y_{rc}(t-\tau)\right]\cdot\tau-T_{ir}(t).    
\end{aligned}
\end{equation}

\begin{figure}[htbp]
	\includegraphics[width=.9\linewidth]{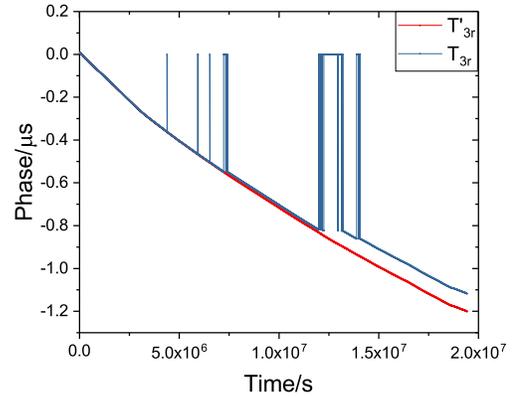}% Here is how to import EPS art
	\caption{\label{disturbance}Measured phase difference ($T_{3r}$) and phase difference substitute ($T'_{3r}$) between clock $H_3$ and the reference clock, $H_4$.}
\end{figure}

At the beginning of the measurement, $a_i(0) = 0$. This design of $a_i(t)$ guarantees that the phase difference data used to generate TSFN are continuous when a disturbed link or clock recovers. The measured phase difference and phase difference substitute between clock $H_3$ and the reference clock $H_4$ is shown in Fig.~\ref{disturbance}. The figure shows that when the fiber link fails and recovers, or when other activities cause a jump in measured phase difference, the substitute remains continuous and smooth, ensuring the stability of the composite.

An adjustment to the weight in Case 2 is made as follows:
\begin{equation}
\begin{aligned}
\omega_i (t)=\min\{\omega_i (t-\tau)+0.001,\omega_i (t)\}. 
\end{aligned}
\end{equation}

This limits the increasing step to 0.001, which is designed to prevent $\omega_i$ from increasing too quickly when a clock is added or recovered. This step, 0.001, was carefully chosen according to experiments. It is small enough that at this growing pace, the composite can remain stable throughout the process. It is also big enough: if all the clocks and transfer links stay in Case 2, variation of any clock's weight between two successive iterations is much smaller than 0.001. In each iteration, after adjusting the weight of each clock, a final normalization step is carried out to ensure that the weights sum up to 1.

\section{\label{5}Results and analysis}

A continuous test over 224 days was performed using this network and algorithm.

\begin{figure}[htbp]
\includegraphics[width=\linewidth]{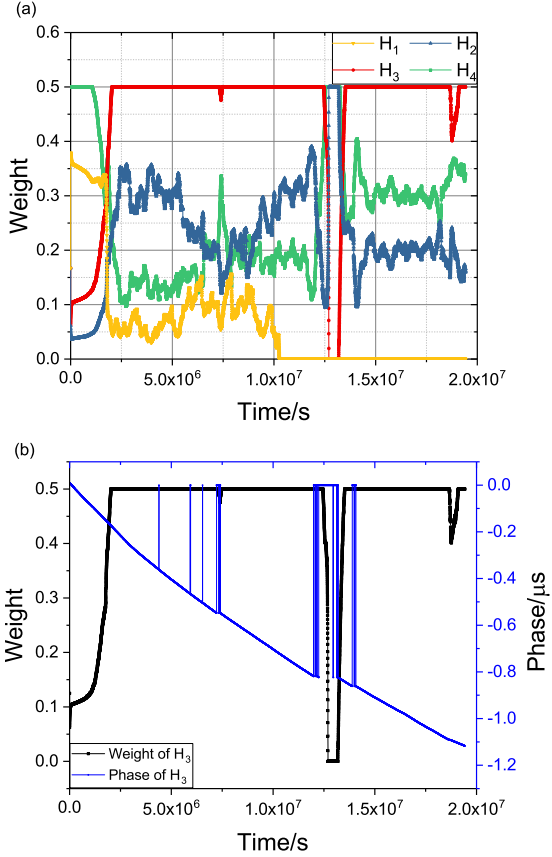}% Here is how to import EPS art
\caption{\label{fig3}(a) Weights of all clocks in the ensemble. (b) Weight and phase (against clock $H_4$) of clock $H_3$.}
\end{figure}

Fig.~\ref{fig3}(a) shows the weight of each clock as a function of time. The initial weights were set such that the reference clock ($H_4$) had a weight of 0.5 and the remaining 0.5 was distributed equally to the other clocks. As the iteration proceeded, it took the algorithm approximately 20 days to train all the parameters, such that the weights came to a steady state at approximately $2.5\times10^6$ s. At that time, it was not the reference clock $H_4$ (green square) but rather clock $H_3$ (red circle) that was given the most weight. Indeed, in our algorithm, the weight of each clock is determined based on its performance, and not influenced by whether or not it is the reference clock.

\begin{figure}[htbp]
	\includegraphics[width=\linewidth]{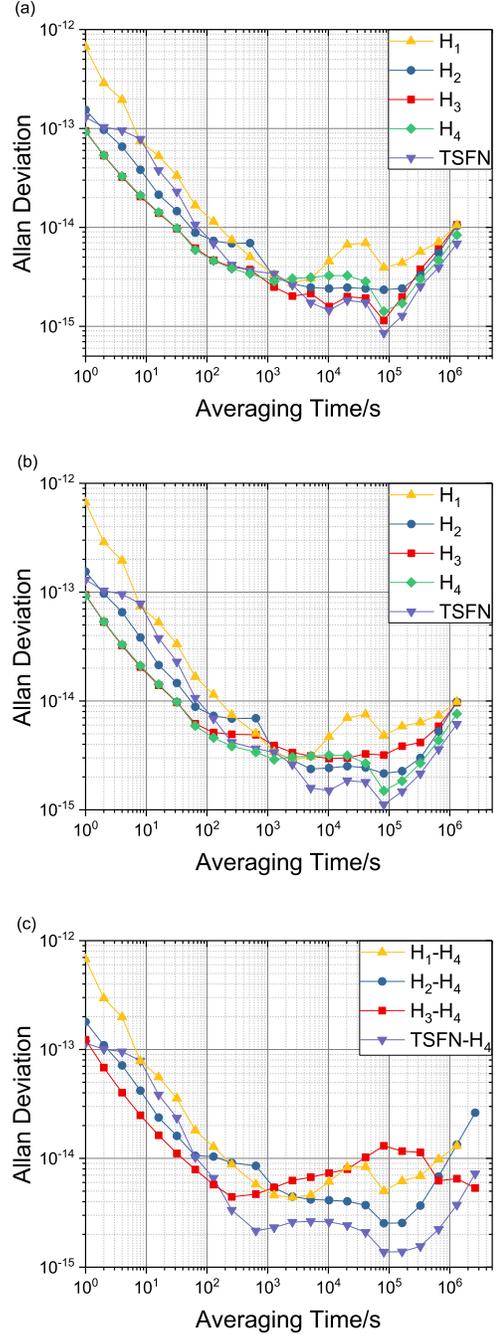}% Here is how to import EPS art
	\caption{\label{fig4} Individual frequency stabilities calculated with the N-Cornered-Hat Algorithm. (a) Individual frequency stabilities of the ensemble and the composite in the period from $2.7\times10^6$ s to $7.2\times10^6$ s (about 52 days). (b) Individual frequency stabilities of the ensemble and the composite in the period from $2.7\times10^6$ s to $1.0\times10^7$ s (about 87 days). (c) Relative frequency stabilities of the ensemble and the composite against the reference clock ($H_4$) in the period from $2.7\times10^6$ s to the end of the test (about 194 days).}
\end{figure}

\begin{figure}[htbp]
	\includegraphics[width=\linewidth]{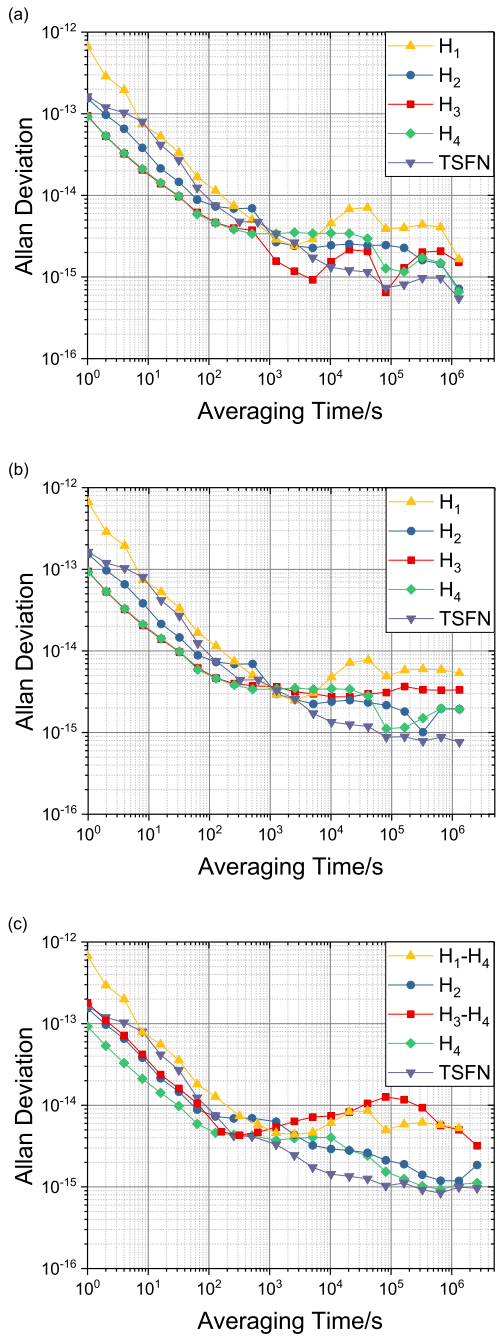}% Here is how to import EPS art
	\caption{\label{method2}Individual frequency stabilities calculated with the Correlated Clock-Ensemble Algorithm. (a) Individual frequency stabilities of the ensemble and the composite in the period from $2.7\times10^6$ s to $7.2\times10^6$ s (about 52 days). (b) Individual frequency stabilities of the ensemble and the composite in the period from $2.7\times10^6$ s to $1.0\times10^7$ s (about 87 days). (c) Individual frequency stabilities of clocks $H_2$, $H_4$, and the composite, as well as relative frequency stabilities of clocks $H_1$ and $H_3$ against the reference clock ($H_4$) in the period from $2.7\times10^6$ s to the end of the test (about 194 days).}
\end{figure}

Taking clock $H_3$ as an example, Fig.~\ref{fig3}(b) shows how its weight changed with phase. The phase data were measured against the reference clock. A fiber link failure appeared to happen at approximately $1.2\times10^7$ s, leading to its weight gradually dropping from 0.5 to 0 after that time. Fortunately, the frequency transfer system recovered soon after the failure, at approximately $1.3\times10^7$ s. Then, the weight of $H_3$ increased to 0.5 again. At other times, there were also some disturbances on $H_3$ because of various reasons. Because of the filters in the algorithm, short-term and small-scale disturbances are filtered out so that they have tiny influence on the weight distribution. In Fig.~\ref{fig3}(b), there are only two visible disturbance-caused effects on the weight of $H_3$, at $7.5\times10^6$ s and $1.85\times10^7$ s. Each time it recovers to 0.5 again, soon after the disturbance stops.

%\begin{figure*}[htbp]
%\includegraphics[width=\textwidth]{fig_4}% Here is how to import EPS art
%\caption{\label{fig4}(a) Individual frequency stabilities of the ensemble and the composite in the period from $2.7\times10^6$s to $7.2\times10^6$s (about 52 days). (b) Individual frequency stabilities of the ensemble and the composite in the period from $2.7\times10^6$s to $1.0\times10^7$s (about 87days). (c) Relative frequency stabilities of the ensemble and the composite against the reference clock ($H_4$) in the period from $2.7\times10^6$s to the end of the test (about 194 days).}
%\end{figure*}

To analyze the stability of TSFN, we used two methods to obtain the individual stabilities of each clock in the ensemble and the composite clock. The individual stabilities shown in Fig.~\ref{fig4} were calculated following the “N-Cornered-Hat Algorithm” discussed in~\cite{40,41}. These stabilities include the frequency drifts. In Fig.~\ref{method2}, the individual stability of each clock in the ensemble was also calculated with the N-Cornered-Hat Algorithm, whereas the stability of the composite clock was obtained following the “Correlated Clock-Ensemble Algorithm” introduced in~\cite{42}. The stabilities in Fig.~\ref{method2} all have the drifts removed.

%The individual frequency stabilities of the composite clock (TSFN) as well as each clock in the ensemble are shown in Fig.~\ref{fig4} in terms of Allan deviation. These individual stabilities are calculated following the algorithm introduced in ~\cite{40,41}. This algorithm does not require the clocks involved in the calculation to be uncorrelated.

 Fig.~\ref{fig4}(a) and Fig.~\ref{method2}(a) show the frequency stabilities in the period from $2.7\times10^6$ s to $7.2\times10^6$ s (about 52 days). In this period, the system was in a steady state before any disturbance occurred on any fiber link. Comparing the stability and weight of each clock, we find that the system decided the weight of each clock based mainly on its stability at the averaging time of around $10^4$ s. At this averaging time, the stabilities of each clock were ordered from highest to lowest as follows: $H_3$, $H_2$, $H_4$, and $H_1$. Correspondingly, the weights of the four clocks in the system (shown in Fig.~\ref{fig3}(a)) ranked in the following order: $\omega_3(t)>\omega_2(t)>\omega_4(t)>\omega_1(t)$. Because the iteration interval $\tau$ in the system is 1200 s, which is also the adjusting interval of the AOG, the stability of TSFN (purple triangle) deteriorated owing to the adjustment before the averaging time of $\tau$. After an averaging time of about $10^4$ s, however, the individual stability of the composite clock was better than all clocks involved.

Fig.~\ref{fig4}(b) and Fig.~\ref{method2}(b) show the frequency stabilities in the period from $2.7\times10^6$ s to $1.0\times10^7$ s (about 87 days). This was before the signal of $H_1$ was lost. During this period, several disturbances occurred with the link of $H_3$. Thus, the stability of $H_3$ deteriorated somewhat compared with its stability in the first period. However, the stability of TSFN did not deteriorate as much as $H_3$. Indeed, it remained better than any of the member clocks. Another point to notice is that, even though in Fig.~\ref{fig4}(b) and Fig.~\ref{method2}(b) the stability of $H_3$ is worse than that of $H_2$ at averaging time of about $10^4$ s, it was still given the most weight throughout the entire period. This is because the Allan deviation was averaged over the entire 87 days of the test, whereas the weight was the exponential weighted average over the previous 2 days. Thus, a disturbance of the frequency transfer system can only affect the weight for a reasonably short period. This ensures that the clock can still play a role in the composite after it recovers.

\begin{table*}[htbp]
	\label{tt}
	\caption{
		Frequency stabilities of each clock with respect to TAI
	}
	\begin{ruledtabular}
		\begin{tabular}{cccccc}
			Clock& $H_{1}$ & $H_{2}$ & $H_{3}$ & $H_{4}$ & TSFN\\
			\hline\\
			Frequency stability\footnote{Frequency stabilities of each clock with respect to TAI at averaging time of 1 month, calculated with experiment data in the period from $2.7\times10^6$ s to the end of the test (about 194 days).} & $3.00\times 10^{-14}$ & $2.81\times 10^{-14}$ & $6.62\times 10^{-15}$ & $4.33\times 10^{-15}$ & $9.39\times 10^{-15}$\\
		\end{tabular}
	\end{ruledtabular}
\end{table*}

Fig.~\ref{fig4}(c) and Fig.~\ref{method2}(c) show the frequency stabilities from $2.7\times10^6$ s to the end of the test (about 194 days). Because clock $H_1$ was lost after approximately $1.0\times10^7$ s and clock $H_3$ experienced a disturbance from $1.2\times10^7$ s to $1.3\times10^7$ s, the remaining data are not sufficient to calculate the individual stability. Instead, in Fig.~\ref{fig4}(c), we show the relative stability of each clock against the reference clock ($H_4$). In Fig.~\ref{method2}(c) we show the individual stability of clocks $H_2$, $H_4$, and the TSFN, calculated using the Correlated Clock-Ensemble Algorithm, as well as the relative stabilities of clock $H_1$ and $H_3$ against the reference clock, $H_4$. For clock $H_1$, the stability is derived from the first $1.0\times10^7$ s. As a result, its data is not sufficient to show the stability at $2.6\times10^6$ s, as other clocks do. The results of this long period also indicate the continuous and stable performance of the composite clock. The stability of the TSFN (purple triangle) was better than any of the member clocks at averaging times larger than 1000 s, even though the stability of clock $H_3$ deteriorated considerably over the course of this long-term analysis. 

Another common method to assess the performance of a free-running time scale is to compare it with an external frequency reference with better frequency stability, such as a cesium fountain or TAI data. However, there is no continuous precise comparison link between TSFN system and external references supporting assessment at short averaging times discussed in this manuscript. With data in the rTAI from BIPM~\cite{43}, we obtained relative frequencies of clocks in the TSFN system with regard to TAI with time interval of about 1 month. Thus, relative frequency stabilities of each clock with regard to TAI at the averaging time of 1 month were estimated as shown in Table~\uppercase\expandafter{\romannumeral1}. The stabilities were calculated with experiment data in the period from $2.7\times10^6$ s to the end of the test (about 194 days) and they are with frequency drifts. Although these results have relatively large error bars because of the short length of data, it may, to some extent, act as a proof that the TSFN performs within the expected range. With more frequent comparison between TSFN system and TAI in the future, parameters in the TSFN algorithm can be set more accurately and the performance of TSFN may be improved.

\section{\label{6}Conclusion}

We generated a real-time free-running time scale based on four remote hydrogen masers scattered across Beijing and connected by urban fiber links. The time scale algorithm is a weighted average of all clocks calculated recursively every 1200 s. The distributed clock ensemble prevents environment-induced clock correlation from influencing the time scale, which is problematic in co-located time scale systems. A disturbance-resistant algorithm was designed to ensure that the proposed TSFN remained continuous and stable in cases of fiber link disturbances or ensemble member changes. A continuous experiment over 224 days was performed. The individual stabilities of each clock in the ensemble as well as that of the composite clock were obtained. The results indicate that the composite clock was more stable than any individual clock in the ensemble for averaging times larger than approximately $10^4$ s. This work contributes to the enrichment of Quadrant \uppercase\expandafter{\romannumeral1}-type clocks (see Fig.~\ref{fig1}). It provides a solution to the long-standing tension between real-time performance and stability/reliability in time scales. Longer tests will be conducted in the future to study the long-term performance of the TSFN.

\begin{acknowledgments}
This work was supported by the National Natural Science Foundation under Grants 91836301 and the National Key Project of Research and Development (No. 2016YFA0302102).
\end{acknowledgments}

% The \nocite command causes all entries in a bibliography to be printed out
% whether or not they are actually referenced in the text. This is appropriate
% for the sample file to show the different styles of references, but authors
% most likely will not want to use it.
\nocite{*}

%\bibliography{aipsamp}% Produces the bibliography via BibTeX.
\bibliographystyle{unsrt}

\end{document}